\documentclass[11pt]{article}
\usepackage{latexsym}

\usepackage{epsfig}
\usepackage{graphicx}

\catcode`\@=11

\@addtoreset{equation}{section}

\def\@normalsize{\@setsize\normalsize{15pt}\xiipt\@xiipt
\abovedisplayskip 14pt plus3pt minus3pt%
\belowdisplayskip \abovedisplayskip
\abovedisplayshortskip  \z@ plus3pt%
\belowdisplayshortskip  7pt plus3.5pt minus0pt}
\def\small{\@setsize\small{13.6pt}\xipt\@xipt
\abovedisplayskip 13pt plus3pt minus3pt%
\belowdisplayskip \abovedisplayskip
\abovedisplayshortskip  \z@ plus3pt%
\belowdisplayshortskip  7pt plus3.5pt minus0pt
\def\@listi{\parsep 4.5pt plus 2pt minus 1pt
            \itemsep \parsep
            \topsep 9pt plus 3pt minus 3pt}}

\def\underline#1{\relax\ifmmode\@@underline#1\else
        $\@@underline{\hbox{#1}}$\relax\fi}
\@twosidetrue \relax

\catcode`@=12

\catcode`\@=11

\def\section{\@startsection{section}{1}{\z@}{3.5ex plus 1ex minus
   .2ex}{2.3ex plus .2ex}{\large\bf}}

\def\ps@headings{\def\@oddfoot{}\def\@evenfoot{}
\def\@oddhead{\hbox{}\hfill
        \makebox[.5\textwidth]{\raggedright\ignorespaces --\thepage{}--
        \hfill }}
\def\@evenhead{\@oddhead}
\def\subsectionmark##1{\markboth{##1}{}}
}

\ps@headings

\catcode`\@=12

\begin{document}

\begin{titlepage}
\begin{flushright}
hep-th/0307144
\end{flushright}
\begin{centering}
\vspace{.8in}
%

%Titlos, Onoma kai Imerominia
%\title{\textbf{Probing the Holography of Near-Horizon $AdS_{5}\times S^{5}$ Geometry}}
%\author{\textbf{E. Papantonopoulos\footnote{lpapa@central.ntua.gr} \hspace{4pt}
%and \hspace{4pt} V. Zamarias\footnote{zamarias@central.ntua.gr}}
%\\ \\ Physics Department, National
%Technical University of Athens,\\ Zografou Campus GR 157 73,
%Athens, Greece} \maketitle

{\large {\bf Probing the Holography of Near-Horizon
$\mathbf{AdS_{5}\times S^{5}}$ Geometry}}
\\

\vspace{.5in} {\bf  E. Papantonopoulos$^{*}$ and  V.
Zamarias$^{**}$}

\vspace{0.3in}

Department of Physics, National Technical University of Athens,\\
Zografou Campus GR 157 73, Athens, Greece\\
\end{centering}

%\vspace{1in}

\begin{abstract} Using the AdS/CFT correspondence we study the holographic principle
 and the CFT/FRW
relations in the near-horizon $AdS_{5}\times S^{5}$ geometry with
a probe D3-brane playing the r$\hat{o}$le of the boundary to this
space. The motion of the probe D3-brane in the bulk, induces a
cosmological evolution on the brane. As the brane crosses the
horizon of the bulk Schwarzschild-AdS$_{5}$ black hole, it probes
the holography of the dual CFT. We test the holographic principle
and we find corrections to CFT/FRW relations in various physical
cases: for radially moving, spinning and electrically charged
D3-brane and for a NS/NS B-field in the bulk.
 \end{abstract}

\begin{flushleft}

 \vspace{.4in} $^{*}$lpapa@central.ntua.gr \\
$^{**}$zamarias@central.ntua.gr

\end{flushleft}
\end{titlepage}

%*\pagebreak \rule{360pt}{2pt}

%\tableofcontents
%\begin{flushleft}
%\rule{360pt}{2pt}
%\end{flushleft}
%\vspace{50pt}

%Proto Kefalaio me Onoma: Introduction
\section{Introduction}
The D-brane solutions of type II supergravity \cite{horow} and
their near-horizon geometry \cite{gibb} are the central
geometrical objects upon which the AdS/CFT correspondence is build
\cite{malda,kleb,witten}. In particular, according to Maldacena
conjecture \cite{malda}, the (3+1)-dimensional world-volume of N
coinciding extremal D3-branes, which give rise to $\mathcal{ N}=4$
supersymmetric SU(N) Yang-Mills (SYM) theory, in the large N
limit, is dual to type IIB superstrings, propagating on the
near-horizon $AdS_{5}\times S^{5}$ background geometry. In a
further proposal \cite{witten2} the thermodynamics of large N,
$\mathcal{ N}=4$ supersymmetric SU(N) Yang-Mills theory is linked
with the thermodynamics of Schwarzschild black holes embedded in
the AdS space \cite{page}. This allows to relate, the
Bekenstein-Hawking entropy, in the Maldacena limit, to the entropy
of Yang-Mills gas at N$\longrightarrow \infty$ and large 't Hooft
coupling $g^{2}_{YM}$N.

Many ideas about AdS/CFT correspondence were influenced by the
intriguing concept of "holography"\cite{hooft,sussk}. The
underlying principle, which was originated in the Bekenstein bound
\cite{bekens}, is based on the notion that the maximal entropy
that can be stored within a given volume will be determined by the
largest black hole fitting inside that volume. Since the entropy
of a black hole is essentially given by its surface area, it
follows directly that all the relevant degrees of freedom of any
system must in some sense live on the boundary enclosing that
system.

The holographic principle imposes on generic field theories
coupled to gravity "holographic area bounds" which limit the
number of physical degrees of freedom. These  "holographic area
bounds" were elegantly applied to cosmology \cite{fischler} where
it was shown that the entropy which crosses the lightlike boundary
of an observable region of the universe, the particle horizon,
should not exceed the horizon area in Planck units.

Recently, using the holographic principle, the entropy bounds in a
radiation dominated closed Friedmann-Robertson-Walker universe was
analyzed \cite{verlide}. It was found a surprising similarity
between Cardy's entropy formula for 1+1 dimensional CFT and the
Friedmann equation governing the evolution of the universe. After
a suitable identification, it was shown that actually the Cardy's
formula \cite{cardy} maps to the Friedmann equation. In a further
development \cite{savonije} this correspondence between Cardy's
formula and the Friedmann equation was tested in the
Randall-Sundrum type model \cite{randall}. In the case where the
bulk is a Schwarschild-AdS background and there is no matter on
the brane, the correspondence between Cardy's formula and the
Friedmann equation is recovered when the brane crosses the black
hole horizon.

Motivated by the work in \cite{verlide,savonije} the cosmological
holographic principle was applied to various black hole
backgrounds under additional physical assumptions. Generalizations
to include a non-vanishing cosmological constant on the brane with
AdS or dS background were studied in
\cite{abdalla,nojiri,zhang,petkou,padilla, youm,cai1,nojiri3}. The
case of having stiff matter on the brane was studied in
\cite{biswas,myung2}. Extensions to charged black hole background
were analyzed in \cite{ohta,cai2,zanon,setare1}, to Kerr-Newmann
rotated black holes in \cite{klemm,setare,jing}. Extension to
Gauss-Bonnet background geometry was considered in
\cite{cai,lidsey,cai3,gregory}, while the holography of
AdS-Taub-Bolt spacetimes was studied in \cite{kakhtari}.

In this work we come back to the original ideas developed in
\cite{savonije} and we study in a systematic way the cosmological
holographic principle as this is expressed through the AdS/CFT
correspondence and the CFT/FRW-cosmologies relations in a generic
static spherically symmetric background geometry with a boundary
simulated by a probe D3-brane moving in this background. An
important issue is that these backgrounds are consistent vacuum
solutions of ten-dimensional string theory, like the
$AdS_{5}\times S^{5}$ background geometry and its near-horizon
limit. Our approach will enable us to test the AdS/CFT
correspondence and the CFT/FRW-cosmologies relations and find
corrections to these relations in various physical situations,
like a spinning or electrically charged D3-probe brane moving in a
near-horizon $AdS_{5}\times S^{5}$ background with or without a
NS/NS B-field.

The necessary machinery for such an investigation has already been
developed in \cite{kehagias} where it was shown that the motion of
the probe D3-brane in this generic background induces on the brane
a cosmological evolution \cite{kraus}, by generating on the brane
an effective energy density and an effective pressure. However,
the induced equation of state on the brane corresponds in most of
the cases, depending on the energy and angular momentum of the
probe D3-brane, to "Mirage" or stiff matter with $|w|>1/3$.
Nevertheless, by choosing particular spherically symmetric
backgrounds sensible cosmological evolution can be generated on
the brane and also brane inflation \cite{pappa,kim,youm2} and exit
from it \cite{papa}.

The paper is organized as follows. After the introduction in
section one, we review the main results of Mirage cosmology needed
to our work, in section two. In section three we study the
holography of near-horizon $AdS_{5}\times S^{5}$ geometry as it is
probed by the D3-brane in various physical cases: the probe
D3-brane  moving in this particular background radially, spinning
or being electrically charged. In section four we study the more
general problem of a probe D3-brane moving in the field of other
D3-black branes. After reviewing the relevant formalism, we study
the holography of this geometry introducing also an NS/NS B-field
in the background. Finally in section five are our conclusions.

%Deftero Kefalaio me Onoma: Mirage Cosmology
\section{A Probe D3-Brane Moving in a Static Spherically Symmetric Background }

 We will consider a probe D3-brane moving in a generic
 static, spherically symmetric background.
 We assume the brane to be light compared to the background so
 that we will neglect the back-reaction.
The background metric we consider has the general form
\begin{equation}\label{in.met}
ds^{2}_{10}=g_{00}(r)dt^{2}+g(r)(d\vec{x})^{2}+
  g_{rr}(r)dr^{2}+g_{S}(r)d\Omega_{5},
\end{equation}
 where $g_{00}$ is negative, and there is also a dilaton field $\Phi$ as well as a $RR$
 background~$C(r)=C_{0...3}(r)$ with a self-dual field strength.

The dynamics on the brane will be governed by the
 Dirac-Born-Infeld action  given by
\begin{eqnarray}\label{B.I. action}
  S&=&T_{3}~\int~d^{4}\xi
  e^{-\Phi}\sqrt{-det(\hat{G}_{\alpha\beta}+(2\pi\alpha')F_{\alpha\beta}-
  B_{\alpha\beta})}
  \\  \nonumber
  &+&T_{3}~\int~d^{4}\xi\hat{C}_{4}+anomaly~terms.
\end{eqnarray}
 The induced metric on the brane is
\begin{equation}\label{ind.metric}
  \hat{G}_{\alpha\beta}=G_{\mu\nu}\frac{\partial x^{\mu}\partial x^{\nu}}
  {\partial\xi^{\alpha}\partial\xi^{\beta}},
\end{equation}
 with similar expressions for $F_{\alpha\beta}$ and
 $B_{\alpha\beta}$.
For an observer on the brane the Dirac-Born-Infeld action is the
volume of the brane trajectory modified by the presence of the
anti-symmetric two-form $ B_{\alpha\beta}$, and worldvolume
anti-symmetric gauge fields $ F_{\alpha\beta}$. This means that,
if there is for example radiation on the D3-brane,
$F_{\alpha\beta}\neq 0$, the brane dynamics will be altered
relative to the case of a brane with no radiation.
 As the brane moves
 the induced world-volume metric becomes a function of
 time, so there is a cosmological evolution from the brane point
 of view \cite{kehagias}.

  In the static
 gauge, $x^{\alpha}=\xi^{\alpha},\alpha=0,1,2,3 $
 using (\ref{ind.metric}) we can calculate the bosonic part of the
 brane Lagrangian which reads
\begin{equation}\label{brane Lagr}
\mathcal{L}=\sqrt{A(r)-B(r)\dot{r}^{2}-D(r)h_{ij}\dot{\varphi}^{i}\dot{\varphi}^{j}}
-C(r),
\end{equation}
where $h_{ij}d \varphi ^{i} d \varphi^{j}$ is the line
 element of the unit five-sphere, and
\begin{equation}\label{met.fun}
  A(r)=g^{3}(r)|g_{00}(r)|e^{-2\Phi},
  B(r)=g^{3}(r)g_{rr}(r)e^{-2\Phi},
  D(r)=g^{3}(r)g_{S}(r)e^{-2\Phi},
\end{equation}
and $C(r)$ is the $RR$ background. The problem is effectively
one-dimensional and can be solved easily. The momenta are given by
\begin{eqnarray}
p_{r}&=&-\frac{B(r)\dot{r}}{\sqrt{A(r)-B(r)\dot{r}^{2}}},\nonumber
\\
p_{i}&=&-\frac{D(r)h_{ij}\dot{\phi}^{j}}{{\sqrt{A(r)-B(r)\dot{r}^{2}-
D(r)h_{ij}\dot{\phi}^{i}\dot{\phi}^{j}}}}. \end{eqnarray} Since
(\ref{brane Lagr}) is not explicitly time dependent and the
$\phi$-dependence is confined to the kinetic term for
$\dot{\phi}$, for an observer in the bulk, the brane moves in a
geodesic parametrised by a conserved energy $E$ and a conserved
angular momentum $l^{2}$ given by
\begin{eqnarray}
E&=&\frac{\partial\mathcal{L}}{\partial \dot{r}}\dot{r}+
\frac{\partial\mathcal{L}}{\partial
\dot{\phi}^{i}}\dot{\phi}^{i}-\mathcal{L}=p_{r}\dot{r}+p_{i}\dot{\phi}^{i}-\mathcal{L},\nonumber
\\ l^{2}&=&h^{ij}\frac{\partial\mathcal{L}}{\partial
\dot{\phi}^{i}} \frac{\partial\mathcal{L}}{\partial
\dot{\phi}^{j}}=h^{ij}p_{i}p_{j}.
\end{eqnarray}
Solving these expressions for $\dot{r}$ and $ \dot{\phi}$ we find
\begin{equation}\label{functions}
\dot{r}^{2}=\frac{A}{B}(1-\frac{A}{(C+E)^{2}}\frac{D+\ell^{2}}{D}),
\,\,
h_{ij}\dot{\varphi}^{i}\dot{\varphi}^{j}=\frac{A^{2}\ell^{2}}{D^{2}(C+E)^{2}}.
\end{equation}
The allowed values of $r$ impose the constraint that $C(r)+E\geq
0$. The induced four-dimensional metric
 on the brane, using (\ref{ind.metric}) in the static gauge, is
\begin{equation}\label{fmet}
d\hat{s}^{2}=(g_{00}+g_{rr}\dot{r}^{2}+g_{S}h_{ij}\dot{\varphi}^{i}\dot{\varphi}^{j})dt^{2}
+g(d\vec{x})^{2}.
\end{equation}
In the above relation we substitute $ \dot{r}^{2}$ and $
h_{ij}\dot{\varphi}^{i}\dot{\varphi}^{j} $ from (\ref{functions}),
and using (\ref{met.fun}) and we get
\begin{equation}
d\hat{s}^{2}=-\frac{g_{00}^{2}g^{3}e^{-2\phi}}{(C+E)^{2}}dt^{2}+g(d\vec{x})^{2}.
\end{equation} We can define the cosmic time $\eta$  as
\begin{equation}\label{cosmic}
 d\eta=\frac{|g_{00}|g^{\frac{3}{2}}e^{-\Phi}}{|C+E|}dt,
\end{equation} so the induced metric becomes
\begin{equation}\label{fin.ind.metric}
d\hat{s}^{2}=-d\eta^{2}+g(r(\eta))(d\vec{x})^{2},
\end{equation}

 The induced metric on the brane (\ref{fin.ind.metric}) is the standard form of a flat
expanding universe. The relation (\ref{cosmic}) will play a
central r$\hat{o}$le in the following. It relates the cosmic time
$\eta$, the time an observer on the brane uses, with the time $t$
that it is used by an observer in the bulk. We can derive the
analogue of the four-dimensional Friedmann equations by defining
$g=\alpha^{2}$

\begin{equation}\label{dens}  \Big{(}\frac
{\dot{\alpha}}{\alpha}\Big{)}^{2}=
\frac{(C+E)^{2}g_{S}e^{2\Phi}-|g_{00}|(g_{S}g^{3}+\ell^{2}e^{2\Phi})}
{4|g_{00}|g_{rr}g_{S}g^{3}}\Big{(}\frac{g'}{g}\Big{)}^{2},
\end{equation}
 where the dot stands for derivative with respect to cosmic time
 while the prime stands for derivatives with respect to $r$. The
 right hand side of (\ref{dens}) can be interpreted in terms of an
 effective matter density on the probe brane
 \begin{equation}\label{denseff1} \frac{8\pi G}{3}\rho_{eff}=
\frac{(C+E)^{2}g_{S}e^{2\Phi}-|g_{00}|(g_{S}g^{3}+\ell^{2}e^{2\Phi})}
{4|g_{00}|g_{rr}g_{S}g^{3}}\Big{(}\frac{g'}{g}\Big{)}^{2},
\end{equation} where $G$ is the four-dimensional Newton's
constant. We can also calculate \begin{eqnarray}  \label{dadot}
\frac{\ddot{\alpha}}{\alpha}&=&\Big{(}1+\frac{g}{g'}\frac{\partial}{\partial
r}\Big{)}
\frac{(C+E)^{2}g_{S}e^{2\Phi}-|g_{00}|(g_{S}g^{3}+\ell^{2}e^{2\Phi})}
{4|g_{00}|g_{rr}g_{S}g^{3}}\Big{(}\frac{g'}{g}\Big{)}^{2}\\
\nonumber &=&\Big{[}1+\frac{1}{2}\alpha\frac{\partial} {\partial
\alpha}\Big{]}\frac{8\pi G}{3}\rho_{eff}.
\end{eqnarray}
If we set the above equal to $-\frac{4\pi G
}{3}(\rho_{eff}+3p_{eff})$ we can define the effective pressure
$p_{eff}$.

 Therefore, the motion of a D3-brane on a
general spherically symmetric background had induced on the brane
an energy density and a pressure. Then, the first and second
Friedmann equations can be derived giving a cosmological evolution
of the brane universe in the sense that an observer on the brane
measures a scale factor $\alpha(\eta)$ of the brane-universe
evolution. This scale factor depends on the position of the brane
in the bulk. This cosmological evolution is known as "Mirage
Cosmology" \cite{kehagias}: the cosmological evolution is not due
to energy density on our universe but on the energy content of the
bulk. In the next section we will also describe the case where the
cosmological evolution can be triggered by the motion of a probe
D3-brane moving in the field of other Dp-branes.

The formalism developed so far allows also for the probe D3-brane
to have a non-zero angular momentum. We can assume also that there
is an electric field on the probe D3-brane. In this case the
action for the D3-brane is given by (\ref{B.I. action}) and in the
background metric (\ref{in.met}), the Lagrangian takes the form
\begin{equation}
\mathcal{L}=\sqrt{A-B\dot{r}^{2}-\mathcal{E}^{2}g^{2}}-C,
\end{equation}
where $\mathcal{E}^{2}=2\pi a^{\prime}E_{i}E^{i}$ and
$E_{i}=-\partial_{t}A_{i}(t)$ in the $A_{0}=0$ gauge and $A$ and
$B$ are given by (\ref{met.fun}). The equations of motion for the
electric field are \begin{equation}
\partial_{t}\Big{(}\frac{g^{2}E_{i}}{\sqrt{A-B\dot{r}^{2}-\mathcal{E}^{2}g^{2}}}
\Big{)}=0 \label{maxeq} \end{equation} and one can find
\cite{kehagias}
\begin{equation} \label{elm}
E_{i}=\frac{\mu_{i}}{g} \sqrt{\frac{A-B\dot{r}^{2}}
{\mu^{2}+g^{2}}}, \,\, \mathcal{E}^{2}= \frac{\mu^{2}}{g^{2}}
\frac{ A-B\dot{r}^{2}}{g^{2}+\mu^{2}},
\end{equation}
where $\mu_{i}$ are integration constants and $\mu^{2}=(2\pi
a^{\prime})\mu_{i}\mu^{i}$. In the case $ \dot{r}=0$, $E_{i}$ is
constant as it is required by ordinary Maxwell equations. From
(\ref{maxeq}) we can calculate $\dot{r}$,
\begin{equation}
\dot{r}^{2}=\frac{A}{B}\Big{(}1-\frac{A}{(C+E)^{2}(1+\mu^{2}g^{-2}}\Big{)},
\end{equation} from which we obtain
$\mathcal{E}^{2}$ after substitution in (\ref{elm})
 \begin{equation}
\mathcal{E}^{2}=\mu^{2}\frac{A^{2}}{(C+E)^{2}(g^{2}+\mu^{2})^{2}}.
\label{strenth} \end{equation} The induced metric on the probe
D3-brane turns out to be
\begin{equation}
d\hat{s}^{2}=-\frac{g_{00}^{2}g^{5}e^{-2\phi}}{(C+E)^{2}(\mu^{2}+g^{2})}dt^{2}+g(d\vec{x})^{2},
\end{equation}
and by defining the cosmic time as
 \begin{equation}\label{elecosmic}
 d\eta=\frac{|g_{00}|g^{\frac{5}{2}}e^{-\Phi}}{|C+E|(\mu^{2}+g^{2})^{\frac{1}{2}} }dt,
\end{equation} the induced metric on the brane becomes
\begin{equation} d\hat{s}^{2}
=-d\eta^{2}+g(r(\eta))(d\vec{x})^{2}. \end{equation}
 Then, the Friedmann equations with
an electric field on the probe D3-brane in a radial motion is
\begin{equation}\label{denselectr}  \Big{(}\frac
{\dot{\alpha}}{\alpha}\Big{)}^{2}=
\frac{(C+E)^{2}(1+\mu^{2}g^{-2})-|g_{00}|g^{3}e^{-2\Phi}}
{4|g_{00}|g_{rr}g^{3}e^{-2\Phi}}\Big{(}\frac{g'}{g}\Big{)}^{2}.
\end{equation}
The dominant contribution to the induced energy density from the
electric field, as can be seen from (\ref{strenth}), is of the
order $\mathcal{E}^{2}$.

The induced cosmological evolution of a brane moving in a
Schwarzschild-AdS background was also discussed in
\cite{savonije}. Nevertheless the Mirage Cosmology allows for more
general backgrounds and more general physical requirements on the
bulk-brane system, allowing to test the AdS/CFT correspondence and
the CFT/FRW-cosmologies relations in various cases. The formalism
which we reviewed in this section can be generalized to a curved
D3-probe brane. This will induce a spatial curvature on the
brane-universe.

%Trito Kefalaio me Onoma: Holography and Mirage Cosmology

\section{The Holographic Description of Near-Horizon $\mathbf{AdS_{5}\times S^{5}}$ Geometry }

We will apply the above described formalism first to the
near-horizon geometry $AdS_{5}\times S^{5}$. There are
Schwarzschild-AdS$_{5}$  black hole solutions in this background
with metric
\begin{equation} ds^{2}=\frac{r^{2}}{L^{2}}\Big{(}-f(r)dt^{2}+(d
\vec{x}) ^{2}\Big{)}
+\frac{L^{2}}{r^{2}}\frac{dr^{2}}{f(r)}+L^{2}d \Omega ^{2}_{5},
\label{bhmetric} \end{equation} where $f(r)=1-\Big{(}
\frac{r_{0}}{r}\Big{)}^{4} $. The $RR$ field is given by
$C=C_{0...3}= \Big{[}
\frac{r^{4}}{L^{4}}-\frac{r^{4}_{0}}{2L^{4}}\Big{]}$.

Using (\ref{in.met}) we find in this background
\begin{eqnarray} g_{00}(r)&=&-\frac{r^{2}}{L^{2}} \Big{(}1-\big{(}
\frac{r_{0}}{r} \Big{)}^{4} \Big{)} =-\frac{1}{g_{rr}} \nonumber
\\ g(r)&=&\frac{r^{2}}{L^{2}} \nonumber \\ g_{s}(r)&=& L^{2},
\label{bachground}
\end{eqnarray} and the brane-universe scale factor is $ \alpha=r/L $.
Substituting the above functions to equation (\ref{denseff1}) we
find the analogue of Friedmann equation on the brane which is
\cite{kehagias}
\begin{equation}
H^{2}=\frac{8\pi G}{3}\rho_{eff}=\frac{1}{L^{2}} \Big{[}
\Big{(}1+\frac{1}{\alpha^{4}}\Big{(}E-r_{0}^{4}/2L^{4}\Big{)}
\Big{)}^{2}-\Big{(}1-\Big{(}\frac{r_{0}}{L}\Big{)}^{4}
\frac{1}{\alpha^{4}}\Big{)}
\Big{(}1+\frac{l^{2}}{L^{2}}\frac{1}{\alpha^{6}}\Big{)}
 \Big{]}, \label{friedbh} \end{equation} where
E is a constant of integration of the background field equations,
expressing the conservation of energy, and it is related to the
black hole mass of the background \cite{steer}, while
$r_{0}^{4}/2L^{4}$ is the constant part of the RR field,
expressing essentially electrostatic energy, and it can be
absorbed into the energy $
 \tilde{E}=E-r_{0}^{4}/2L^{4}$. This Friedmann equation
 describes the cosmological evolution of a contracting or
 expanding universe depending on the motion of the probe brane. This
 motion in turn depends on two parameters the energy $\tilde{E}$
 and the angular momentum $l^{2}$. These two parameters specify various
 trajectories of the probe brane. The scale factor $\alpha$ comes
 in various powers, indicating that (\ref{friedbh}) describes the cosmological
 evolution of various
 kind of Mirage or stiff cosmological matter.

\subsection{A Probe D3-brane Moving Radially in a Near-Horizon $\mathbf{AdS_{5}\times S^{5}}$
Black Hole Background}

We will follow first the motion of the probe D3-brane in the case
of $l^{2}=0$. Defining the dimensionless parameter $
a=l^{2}/{L^{2}}\alpha^{6} $, equation (\ref{friedbh}) becomes
\begin{equation} H^{2}=\frac{8\pi G}{3}\rho_{eff}=\frac{1}{L^{2}}
\Big{[} \Big{(}1+\frac{\tilde{E}}{\alpha^{4}}
\Big{)}^{2}-\Big{(}1-\Big{(}\frac{r_{0}}{L}\Big{)}^{4}
\frac{1}{\alpha^{4}}\Big{)} \Big{(}1+a \Big{)}
 \Big{]}. \label{friedbha} \end{equation}

Using equations (\ref{dadot}) and (\ref{friedbha}), the second
Friedmann equation in this background reads
\begin{equation} \label{friedgsec}
\dot{H}=-\frac{2}{L^{2}}
\Big{[}2\frac{\tilde{E}}{\alpha^{4}}\Big{(}
1+\frac{\tilde{E}}{\alpha^{4}}
\Big{)}+\Big{(}\frac{\alpha_{0}}{\alpha}
\Big{)}^{4}\Big{(}1+a\Big{)}-\frac{3}{2}a\Big{(}1-\Big{(}\frac{\alpha_{0}}{\alpha}
\Big{)}^{4}\Big{)} \Big{]},
\end{equation} and the effective pressure, using again (\ref{dadot}), is
\begin{equation}
p_{eff}=\frac{1}{8 \pi GL^{2}}
\Big{[}\Big{(}\frac{\alpha_{0}}{\alpha}
\Big{)}^{4}+5\frac{\tilde{E}^{2}}{\alpha^{8}}
+2\frac{\tilde{E}}{\alpha^{4}} +7a\Big{(}\frac{\alpha_{0}}{\alpha}
\Big{)}^{4}-3a    \Big{]}. \label{effpres}
\end{equation} From (\ref{friedbha}) we also have the
effective energy density
\begin{equation}
\rho_{eff}=\frac{3}{8 \pi GL^{2}}
\Big{[}\Big{(}\frac{\alpha_{0}}{\alpha}
\Big{)}^{4}+\frac{\tilde{E}^{2}}{\alpha^{8}}
+2\frac{\tilde{E}}{\alpha^{4}}-a\Big{(}
1-\Big{(}\frac{\alpha_{0}}{\alpha} \Big{)}^{4}\Big{)} \Big{]}.
\label{effnerg}
\end{equation}
It is instructive to consider the equation of state
$p_{eff}=w\rho_{eff}$ where $w$ is given by
\begin{equation}
w=\frac{1}{3}\Big{[}\frac{\Big{(}\frac{\alpha_{0}}{\alpha}
\Big{)}^{4}+5\frac{\tilde{E}^{2}}{\alpha^{8}}
+2\frac{\tilde{E}}{\alpha^{4}} +7a\Big{(}\frac{\alpha_{0}}{\alpha}
\Big{)}^{4}-3a }{\Big{(}\frac{\alpha_{0}}{\alpha}
\Big{)}^{4}+\frac{\tilde{E}^{2}}{\alpha^{8}}
+2\frac{\tilde{E}}{\alpha^{4}}-a\Big{(}
1-\Big{(}\frac{\alpha_{0}}{\alpha} \Big{)}^{4}\Big{)}}\Big{]}.
\label{ww}
\end{equation}
As the brane moves in the Schwarzschild-AdS$_{5}$  black hole
background, the equation of state is parametrized by the energy of
the bulk and the angular momentum of the brane. However, when
$\tilde{E}=0$ and $a=0$ the brane-universe is radiation dominated
at any position in the bulk, as can be seen from (\ref{ww}). This
is expected, because the only scale in the theory is the energy
scale $\title{E}$ and putting it to zero the theory is scale
invariant, while a non-zero angular momentum induces on the brane
all kind of Mirage matter. For $\tilde{E}=0$ and $a=0$ the
equations (\ref{friedbha}), (\ref{friedgsec}), (\ref{effpres}) and
(\ref{effnerg}) become
\begin{eqnarray}
H^{2}&=& \frac{1}{L^{2}}\Big{(} \frac{\alpha_{0}}{\alpha}\Big{)}
^{4} \label{radiauniff} \\ \dot{H}&=& -\frac{2}{L^{2}}\Big{(}
\frac{\alpha_{0}}{\alpha}\Big{)} ^{4} \label{radiauniss} \\
\rho_{eff}&=& \frac{3}{8
\pi G L^{2}}\Big{(} \frac{\alpha_{0}}{\alpha}\Big{)} ^{4} \label{radiaunirho} \\
p_{eff}&=& \frac{1}{8 \pi G L^{2}}\Big{(}
\frac{\alpha_{0}}{\alpha}\Big{)} ^{4} \label{radiaunip}
\end{eqnarray}

In \cite{verlide} it was shown that for a  radiation dominated
universe, the first and second Friedmann equations can be written
in a way similar to Cardy and Smarr formulae respectively. This in
turn means that the Cardy and Smarr formulae can be expressed in
terms of cosmological quantities of a radiation dominated
universe. The Hubble entropy is defined by
\begin{equation} S_{H}=HV/2G,
\end{equation} while the Bekenstein-Hawking energy is $E_{BH}=3V/4 \pi G
r^{2}$. We can also define the Hubble temperature $T_{Hubble}$
\footnote{where the minus sign is necessary to get a positive
result, since in a radiation dominated universe the expansion
always slows down. Further, to avoid the danger of dividing by
zero, we assume that we are in a strongly self-gravitating phase
with $H r \geq 1$.} from
\begin{equation} T_{Hubble} \equiv -\frac{\dot{H}}{2\pi H}.
\label{temphubble}
\end{equation} Using the above definitions, we can verify that the first Friedmann equation
(\ref{radiauniff}) can be written as the Cardy-Verlinde formula
\begin{equation} S_{H} = \frac{2\pi
r}{3}\sqrt{E_{BH}(2E-kE_{BH})}\,\,, \label{cosmcardy-verl}
\end{equation} where $kE_{BH}=0$, because our brane-universe is flat,
while the second Friedmann equation (\ref{radiauniss}) can be
written as the Smarr formula
\begin{equation} kE_{BH} = 3(E + pV - T_{Hubble}S_{H}).
\label{smarrfor}
\end{equation}

We can also express the Cardy-Verlinde and Smarr formulae in terms
of thermodynamical quantities of the dual CFT theory. The
Bekenstein-Hawking entropy of the AdS black hole is given by the
area of the horizon measured in bulk Planckian units. For
spherically symmetric backgrounds the entropy is defined by
\begin{equation} S=\frac{V_{H}}{4G_{Bulk}},
\end{equation} where $G_{Bulk}$ is the bulk Newton's
constant and $V_{H}$ is the area of the horizon
$V_{H}=r_{H}^{n}Vol(S^{n})$ with $Vol(S^{n})$ the volume of a unit
n-sphere. The total entropy is constant during the evolution, but
the entropy density varies with the time. If we define $s=S/V$,
the entropy density is \begin{equation} \label{entropdens}
s=\Big{(}\frac{r_{0}}{r}\Big{)}^{n}\frac{(n-1)}{4G L},
\end{equation} where we have used the relation $G_{Bulk}=G L/(n-1)$ \cite{gubster}.

An observer in the bulk is using the AdS time t and measures the
Hawking temperature from the equation
\begin{equation} T_{H}=\frac{h^{\prime}(r_{0})}{4\pi},
\label{hawtemp}
\end{equation} where the differentiation is with respect to r, which is the distance
variable in the bulk. In our background
$h(r)=\frac{r^{2}}{L^{2}}\Big{[}
1-\Big{(}\frac{r_{0}}{r}\Big{)}^{4} \Big{]}$, and the Hawking
temperature becomes \begin{equation} T_{H}=\frac{r_{0}}{\pi
L^{2}}. \label{hawkt}
\end{equation}
An observer on the brane is using the cosmic time $\eta$ defined
by (\ref{cosmic}). Using equations (\ref{bachground}) and the fact
that the probe brane is moving in a geodesic where $\tilde{E}=0 $,
equation (\ref{cosmic}) becomes \begin{equation} \label{relcosads}
d\eta=\frac{r}{L}\Big{(}1-\Big{(}\frac{r_{0}}{r}
\Big{)}^{4}\Big{)}dt.
 \end{equation}

  It was argue in
\cite{witten2} that the energy, entropy and temperature of a CFT
at high temperatures can be identified up to a conformal factor
with the mass, entropy and Hawking temperature of the AdS black
hole. To fix the conformal factor, according  to the AdS/CFT
correspondence the CFT lives on a space-time which can be
identified with the asymptotic boundary of the AdS black hole. The
asymptotic form of the metric in our case, using (\ref{bhmetric}),
is \begin{equation} \label{limit} \lim_{r \rightarrow \infty}
\Big{[}\frac{L^{2}}{r^{2}}ds^{2} \Big{]}= \lim_{r \rightarrow
\infty} \Big{[}-f(r)dt^{2}+(d \vec{x})
^{2}+\frac{L^{4}}{r^{4}}\frac{dr^{2}}{f(r)}+\frac{L^{4}}{r^{2}}d
\Omega ^{2}_{5}  \Big{]}.
\end{equation}
Taking the limit of $f(r)$, relation (\ref{limit}) becomes
\begin{equation}\label{limittime}
\lim_{r \rightarrow \infty} \Big{[}\frac{L^{2}}{r^{2}}ds^{2}
\Big{]}=  -dt^{2}+(d \vec{x}) ^{2}+L^{4 }d \Omega ^{2}_{3}.
\end{equation} Therefore the CFT time and the AdS time are related
through the conformal factor  $r/L$ . Note that the same result
can be obtained considering relation ({\ref{relcosads}) for large
r. Both procedures give the same result because our probe brane is
flat as it is discussed in \cite{medved}.

Having fixed the conformal factor we can relate CFT  and the
Hawking temperature
\begin{equation}
T_{CFT}=\frac{1}{\alpha}T_{H}=\frac{L}{r}T_{H}, \label{brcft}
\end{equation} and then the Cardy-Verlinde and Smarr formulae can be derived using the
AdS/CFT correspondence. Using (\ref{entropdens}) for $n=3$ the CFT
entropy density is
\begin{equation} \label{cftentropdens}
s_{CFT}=\frac{ 1}{2 G L}\Big{(}\frac{r_{0}}{r}\Big{)}^{3},
\end{equation} while the Casimir energy is defined by (Smarr
formula)
\begin{equation}
E_{C}=3\Big{(} E + pV-T_{CFT}S_{CFT} \Big{)}.\label{casimir}
\end{equation} Equation (\ref{casimir}) can be written as $
\rho_{C}=3\Big{(} \rho_{eff}+p_{eff}-T_{CFT}s_{CFT}\Big{)}$ from
which after substitution of the relevant quantities we get
$E_{C}=0$ as expected in a flat radiation dominated
brane-universe. Then, the Cardy-Verlinde formula
\begin{equation}
S_{CFT}= \frac{2 \pi r}{3}\sqrt{\frac{3}{2\pi r}S_{C}(2E-E_{C})},
\label{cftcardy-verl}
\end{equation} with $S_{C}=\frac{ V}{2 G
r}\Big{(}\frac{r_{0}}{r}\Big{)}^{2}$ is trivially satisfied.

At the special moment at which the brane crosses the bulk black
hole horizon, the Hubble temperature  (\ref{temphubble}) becomes
$T_{Hubble}=1/\pi L$. On the other hand the CFT temperature
$T_{CFT}$ using (\ref{hawtemp}) and (\ref{brcft}) on the horizon
becomes $T_{CFT}=1/\pi L$, therefore on the horizon
$T_{CFT}=T_{Hubble}$. It is also easy to check that on the horizon
$S_{CFT}=S_{H}$ and $ E_{C}=kE_{BH} $. Therefore as the brane
crosses the black hole horizon equation ({\ref{cftcardy-verl}) is
equivalent to (\ref{cosmcardy-verl}) and equation (\ref{casimir})
is equivalent to (\ref{smarrfor}). This is known as
CFT/FRW-cosmologies correspondence, expressing a special
interrelation between thermodynamical and geometrical quantities
in a radiation dominated universe.

These results are in agreement with various studies of a brane
moving in a Schwarzschild-AdS$_{5}$  black hole background
\cite{savonije,nojiri,myung,wang}. In the next sections we will
test the AdS/CFT correspondence and the CFT/FRW-cosmologies
relations under various physical conditions on the brane and the
bulk.

\subsection{A Spinning Probe D3-Brane Moving in a Near-Horizon $\mathbf{AdS_{5} \times S^{5}}$
 Black Hole Background}

In this section we will study the motion of the probe D3-brane
carrying a non-trivial angular momentum in a near-horizon
$AdS_{5}\times S^{5}$ black hole background. If $a \neq 0 $, we
can see from (\ref{ww}) that the non-zero angular momentum induces
Mirage matter on the brane with $w$ taking any value. We will set
$\tilde{E}=0$ and we will find the corrections of the various
quantities involved due to the angular momentum. Using
(\ref{friedbha}) and (\ref{friedgsec}) for $\tilde{E}=0$, the
Hubble entropy, Bekenstein-Hawking energy and Hubble temperature
become
\begin{eqnarray} S_{H}&=& \frac{V}{2 G L}\Big{[}\Big{(}
\frac{\alpha_{0}}{\alpha}\Big{)} ^{4} -a \Big{(} 1-\Big{(}
\frac{\alpha_{0}}{\alpha}\Big{)} ^{4} \Big{)}\Big{]}^{1/2} \label{ahubble}\\
E_{BH}&=&\frac{3V}{4 \pi G L^{2}
\alpha^{2}} \label{hubenergy} \\
T_{Hubble}&=&\frac{1}{\pi L} \Big{[} \frac{\Big{(}\Big{(}
\frac{\alpha_{0}}{\alpha}\Big{)} ^{4} -\frac{a}{2} \Big{(}
3-5\Big{(} \frac{\alpha_{0}}{\alpha}\Big{)} ^{4}
\Big{)}\Big{)}}{\Big{(}\Big{(} \frac{\alpha_{0}}{\alpha}\Big{)}
^{4} -a \Big{(} 1-\Big{(} \frac{\alpha_{0}}{\alpha}\Big{)} ^{4}
\Big{)}\Big{)}^{1/2}} \Big{]}.  \label{lhubble}
\end{eqnarray}
Using the above relations we can verify that equations
(\ref{cosmcardy-verl}) and ({\ref{smarrfor}) are satisfied,
suggesting that despite the brane-universe is not radiation
dominated, the first and second Friedmann equations can still be
written as the cosmological Cardy-Verlinde and Smarr formulae
respectively, having the additional information of the spinning
probe brane.

The angular momentum does not appear in the definition of the
cosmic time (\ref{cosmic}). On the other hand the asymptotic limit
of the metric is the same as in the case of $a=0$, because the
bulk metric in independent of the angular momentum of the brane.
Therefore, the conformal factor is the same $r/L$ as before and
the CFT and Hawking temperatures are related the same way as in
(\ref{brcft}). Hence, the presence of a non-trivial angular
momentum on the probe D3-brane maintains the exact AdS/CFT
correspondence.

The Casimir energy, using (\ref{effpres}) and (\ref{effnerg}),
becomes
\begin{equation} E_{C}=\frac{3 V a}{4 \pi G L^{2}}\Big{[}5
\Big{(} \frac{\alpha_{0}}{\alpha}\Big{)} ^{4}-3 \Big{]}
\label{lcasimir}.
\end{equation} It is proportional to the angular momentum parameter $a$,
while the CFT entropy is $S_{CFT} =Vs_{CFT}$ where $s_{CFT}$ is
given by (\ref{cftentropdens}). At the moment the probe D3-brane
crosses the bulk black hole horizon, the Hubble temperature from
equation (\ref{lhubble}) becomes $T_{Hubble}=\frac{1}{\pi L}(1+a)$
while $T_{CFT}=1/\pi L$ as before. Therefore at the horizon
$T_{Hubble}\neq T_{CFT}$. One can easily check, using
(\ref{ahubble}) and (\ref{lcasimir}), that at the horizon
$\alpha=\alpha_{0}$, we have
\begin{eqnarray} S_{H}&=& S_{CFT}\\ E_{C}&\neq& kE_{BH}.
\end{eqnarray} Therefore, as the spinning probe brane crosses the
black hole horizon the CFT/FRW-cosmologies relations break down.

\subsection{Electric Field on the Probe D3-brane}

In this section we will consider an electric field on the probe
D3-brane. To simplify the discussion we will assume $l^{2}=0$. As
we can see from (\ref{denselectr}) the electric field introduces a
radiation term on the probe brane. Substituting the functions
(\ref{bachground}) into (\ref{denselectr}) and defining
$\tilde{\mathcal{E}}=\mu^{2}$, the first Friedmann equation
becomes
\begin{equation}
H^{2}=\frac{8 \pi G}{3} \rho_{eff}+\frac{8 \pi G}{3}
\rho_{rad}=\frac{1}{L^{2}} \Big{[}
\Big{(}1+\frac{\tilde{E}}{\alpha^{4}}\Big{)}^{2}
-\Big{(}1-\Big{(}\frac{\alpha_{0}}{\alpha}\Big{)}^{4}\Big{)}\Big{]}
+\frac{1}{L^{2}}\Big{(}1+\frac{\tilde{E}}{\alpha^{4}}\Big{)}^{2}\frac{\tilde{\mathcal{E}}}{\alpha^{4}},
 \label{effriedbh} \end{equation}
 while the second Friedmann equation can be easily calculated
 \begin{eqnarray} \dot{H}&=&-\frac{2}{L^{2}} \Big{[}2
 \frac{\tilde{E}}{\alpha^{4}}
\Big{(}1+\frac{\tilde{E}}{\alpha^{4}}\Big{)}
+\Big{(}\frac{\alpha_{0}}{\alpha}\Big{)}^{4}
+\frac{\tilde{\mathcal{E}}}{\alpha^{4}}
\Big{(}1+\frac{\tilde{E}}{\alpha^{4}}\Big{)}\Big{(}1+3\frac{\tilde{E}}{\alpha^{4}}\Big{)}
 \Big{]}.
\label{esfried}
\end{eqnarray}
From (\ref{esfried}) using (\ref{dadot}) the pressure can be
calculated \begin{equation} p=p_{eff}+p_{rad}=\frac{1}{8 \pi G
L^{2}}\Big{[}\Big{(}\frac{\alpha_{0}}{\alpha}\Big{)}^{4}+\frac{\tilde{E}}{\alpha^{4}}\Big{(}
5\frac{\tilde{E}}{\alpha^{4}}+2 \Big{)} \Big{]}+\frac{1}{8 \pi G
L^{2}}\Big{[}\frac{\tilde{\mathcal{E}}}{\alpha^{4}}\Big{(}
1+10\frac{\tilde{E}}{\alpha^{4}}+9\frac{\tilde{E}^{2}}{\alpha^{8}}\Big{)}
\Big{]}.
\end{equation}
Demanding to have a radiation dominated brane-universe ($w=1/3$),
we get two solutions for the parameters $\tilde{E}$ and
$\tilde{\mathcal{E}}$
\begin{eqnarray}
\tilde{E}&=& 0\\  \tilde{E}&=&-\alpha^{4}
\frac{2\tilde{\mathcal{E}} /
\alpha^{4}}{1+2\tilde{\mathcal{E}}/\alpha^{4}}. \label{secondsol}
\end{eqnarray}

The parameter $\tilde{\mathcal{E}}$ being proportional to the
energy density of the electric field (relation (\ref{strenth})),
introduces another energy scale in the theory and we expect to
affect the motion of the probe D3-brane. For
$\tilde{\mathcal{E}}=0$ we recover the results of a radially
moving probe D3-brane. We will study first the solution
$\tilde{E}= 0$. The two Friedmann equations and the energy density
and pressure on the brane become
\begin{eqnarray}
H^{2}&=& \frac{1}{L^{2}}\Big{[}\Big{(}
\frac{\alpha_{0}}{\alpha}\Big{)}
^{4}+\frac{\tilde{\mathcal{E}}}{\alpha^{4}}\Big{]}
\label{eradiauniff} \\ \dot{H}&=& -\frac{2}{L^{2}}\Big{[}\Big{(}
\frac{\alpha_{0}}{\alpha}\Big{)}
^{4}+\frac{\tilde{\mathcal{E}}}{\alpha^{4}}\Big{]}
\label{eradiauniss} \\ \rho&=& \frac{3}{8 \pi G
L^{2}}\Big{[}\Big{(} \frac{\alpha_{0}}{\alpha}\Big{)}
^{4}+\frac{\tilde{\mathcal{E}}} {\alpha^{4}}\Big{]}
\label{eradiaunirho} \\ p&=& \frac{1}{8 \pi G L^{2}}\Big{[}\Big{(}
\frac{\alpha_{0}}{\alpha}\Big{)}
^{4}+\frac{\tilde{\mathcal{E}}}{\alpha^{4}}\Big{]}.
\label{eradiaunip}
\end{eqnarray}
The above relations can be represented as corrections to relations
(\ref{radiauniff})-(\ref{radiaunip}), due to the presence of an
electric field on the probe D3-brane. Using
(\ref{eradiauniff})-(\ref{eradiaunip}), the Hubble entropy,
Bekenstein-Hawking energy and Hubble temperature can be calculated
to be
\begin{eqnarray} S_{H}&=& \frac{V}{2 G L}\Big{[}\Big{(}
\frac{\alpha_{0}}{\alpha}\Big{)} ^{4} +\frac{\tilde{\mathcal{E}}}{\alpha^{4}}\Big{]}^{1/2}
 \\ E_{BH}&=&\frac{3V}{4 \pi  G L^{2} \alpha^{2}}\\
T_{Hubble}&=&\frac{1}{\pi L} \Big{[}\Big{(}
\frac{\alpha_{0}}{\alpha}\Big{)} ^{4}
+\frac{\tilde{\mathcal{E}}}{\alpha^{4}}\Big{]}^{1/2}.
\label{elhubble}
\end{eqnarray}
Observe that the Hubble entropy and the Hubble temperature have
been modified by the same extra radiation term. The two Friedmann
equations (\ref{eradiauniff}) and (\ref{eradiauniss}) can be
written as the cosmological Cardy-Verlinde and Smarr formulae
respectively, modified by the radiation term due to the electric
field
\begin{eqnarray}
S_{H} &=& \frac{2\pi
r}{3}\sqrt{E_{BH}\Big{(}2(E+E_{rad})-kE_{BH}\Big{)}}\,\,,
\label{ecosmcardy-verl} \\kE_{BH}& =& 3\Big{(}(E+E_{rad}) +
(p+p_{rad})V - T_{Hubble}S_{H}\Big{)}. \label{esmarrfor}
\end{eqnarray}

To calculate the thermodynamic quantities of the dual theory we
have to find the conformal factor. The conformal time in presence
of the electric field, using (\ref{elecosmic}), becomes
\begin{equation} d\eta=\alpha \Big{(}1-\Big{(}\frac{r_{0}}{r}\Big{)}
^{4}\Big{)}\Big{(}1+\frac{\tilde{\mathcal{E}}}{\alpha^{4}}\Big{)}^{-1/2}dt.
\end{equation} Then the conformal factor is
\begin{equation}\lim_{r\rightarrow\infty}\frac{dt}{d\eta}=\frac{1}{\alpha}
\Big{(}1+\frac{\tilde{\mathcal{E}}}{\alpha^{4}}\Big{)}^{1/2}.
\label{econffac}
\end{equation}
If we calculate the conformal factor using the asymptotic form of
the metric we will find a different result. The reason is that the
bulk metric does not "see" the electric field on the brane. This
case is similar to the case of a brane having a non-zero tension
discussed in the literature \cite{padilla,wang,medved}.

 The CFT and Hawking temperature are now related by
$T_{CFT}=\Big{(}\Big{(}1+\frac{\tilde{\mathcal{E}}}{\alpha^{4}}\Big{)}^{1/2}/\alpha\Big{)}
T_{H} $ and using the Hawking temperature (\ref{hawkt}), which
does not change because it is a bulk quantity, it becomes
\begin{equation}
T_{CFT}=\frac{1}{\pi L}\Big{(}\frac{r_{0}}{r} \Big{)}
\Big{(}1+\frac{\tilde{\mathcal{E}}}{\alpha^{4}}\Big{)}^{1/2}.
\end{equation}
The Casimir energy can be calculated from the first law of
thermodynamics $Tds=d\rho+3(\rho+p-Ts)dr/r$ and we get
\begin{equation}
E_{C}=\frac{3V}{2 \pi G L^{2}}\Big{[}\Big{(}
\frac{\alpha_{0}}{\alpha}\Big{)}
^{4}+\frac{\tilde{\mathcal{E}}}{\alpha^{4}}-\Big{(}\frac{\alpha_{0}}{\alpha}\Big{)}
^{4}\Big{(}1+\frac{\tilde{\mathcal{E}}}{\alpha^{4}}\Big{)}^{1/2}\Big{]}.
\end{equation} Using this expression for the Casimir energy we can
check that the Cardy-Verlinde formula (\ref{cftcardy-verl}) is not
satisfied.

As the brane crosses the bulk black hole horizon, the
AdS/FRW-cosmologies relations break down because $S_{H}\neq
S_{CFT}$ and $ kE_{BH}\neq E_{C}$.

The second solution (\ref{secondsol}) has similar behavior like
the first one. The first and second Friedmann equations become
\begin{eqnarray}
H^{2}&=& \frac{1}{L^{2}}\Big{[}\Big{(}
\frac{\alpha_{0}}{\alpha}\Big{)}
^{4}-\frac{\tilde\mathcal{{E}}}{\alpha^{4}}
\Big{(}3-4\Big{(}\frac{\alpha_{0}}{\alpha}\Big{)}^{4}\Big{)}-4\Big{(}\frac{\tilde\mathcal{{E}}}{\alpha^{4}}\Big{)}^{2}
\Big{(}1-\Big{(}\frac{\alpha_{0}}{\alpha}\Big{)}^{4}
\Big{)}\Big{]}\Big{(}1+2\frac{\tilde\mathcal{{E}}}{\alpha^{4}}\Big{)}^{-2}\\
\dot{H}&=&-\frac{2}{L^{2}}\Big{[}\Big{(}
\frac{\alpha_{0}}{\alpha}\Big{)}
^{4}-\frac{\tilde\mathcal{{E}}}{\alpha^{4}}
\Big{(}3-4\Big{(}\frac{\alpha_{0}}{\alpha}\Big{)}^{4}\Big{)}-4\Big{(}\frac{\tilde\mathcal{{E}}}{\alpha^{4}}\Big{)}^{2}
\Big{(}1-\Big{(}\frac{\alpha_{0}}{\alpha}\Big{)}^{4}
\Big{)}\Big{]}\Big{(}1+2\frac{\tilde\mathcal{{E}}}{\alpha^{4}}\Big{)}^{-2}.
\end{eqnarray}

The Hubble entropy, Bekenstein-Hawking energy and Hubble
temperature for this solution become
\begin{eqnarray} S_{H}&=& \frac{V}{2 G L}\Big{[}\Big{(}
\frac{\alpha_{0}}{\alpha}\Big{)}
^{4}-\frac{\tilde\mathcal{{E}}}{\alpha^{4}}
\Big{(}3-4\Big{(}\frac{\alpha_{0}}{\alpha}\Big{)}^{4}\Big{)}-4\Big{(}\frac{\tilde\mathcal{{E}}}{\alpha^{4}}\Big{)}^{2}
\Big{(}1-\Big{(}\frac{\alpha_{0}}{\alpha}\Big{)}^{4}
\Big{)}\Big{]}^{1/2}\Big{(}1+2\frac{\tilde\mathcal{{E}}}{\alpha^{4}}\Big{)}^{-1}\label{eshubble} \\
 E_{BH}&=&\frac{3V}{4 \pi G L^{2} \alpha^{2}} \label{esenergy} \\
T_{Hubble}&=&\frac{1}{\pi L} \Big{[}\Big{(}
\frac{\alpha_{0}}{\alpha}\Big{)}
^{4}-\frac{\tilde\mathcal{{E}}}{\alpha^{4}}
\Big{(}3-4\Big{(}\frac{\alpha_{0}}{\alpha}\Big{)}^{4}\Big{)}-4\Big{(}\frac{\tilde\mathcal{{E}}}{\alpha^{4}}\Big{)}^{2}
\Big{(}1-\Big{(}\frac{\alpha_{0}}{\alpha}\Big{)}^{4}
\Big{)}\Big{]}^{1/2}\Big{(}1+2\frac{\tilde\mathcal{{E}}}{\alpha^{4}}\Big{)}^{-1}.
\label{esthubble}
\end{eqnarray}

Using (\ref{eshubble})-(\ref{esthubble}) it is easy to verify the
cosmological Cardy-Verlinde and Smarr formulae
(\ref{cosmcardy-verl}) and (\ref{smarrfor}).

The conformal factor for this solution is
\begin{equation}
d\eta=\alpha \Big{(}1-\Big{(}\frac{r_{0}}{r}\Big{)}
^{4}\Big{)}\Big{(}1+2\frac{\tilde{\mathcal{E}}}{\alpha^{4}}\Big{)}
\Big{(}1+\frac{\tilde{\mathcal{E}}}{\alpha^{4}}\Big{)}^{-1/2}dt.
\end{equation}
From which we calculate the conformal factor
\begin{equation}\lim_{r\rightarrow\infty}\frac{dt}{d\eta}=\frac{1}{\alpha}
\Big{(}1+\frac{\tilde{\mathcal{E}}}{\alpha^{4}}\Big{)}^{1/2}
\Big{(}1+2\frac{\tilde{\mathcal{E}}}{\alpha^{4}}\Big{)}^{-1}.
\end{equation}
Using this conformal factor the CFT temperature becomes
\begin{equation}
T_{CFT}=\frac{1}{\pi
L}\Big{(}\frac{r_{0}}{r}\Big{)}\Big{(}1+\frac{\tilde{\mathcal{E}}}{\alpha^{4}}\Big{)}^{1/2}
\Big{(}1+2\frac{\tilde{\mathcal{E}}}{\alpha^{4}}\Big{)}^{-1}.
\end{equation} while the Casimir energy is \begin{eqnarray} E_{C}&=&\frac{3 V}{2\pi G L^{2}}
\Big{(}1+2\frac{\tilde{\mathcal{E}}}{\alpha^{4}}\Big{)}^{-2}
\Big{[}\Big{(} \frac{\alpha_{0}}{\alpha}\Big{)}
^{4}-\frac{\tilde\mathcal{{E}}}{\alpha^{4}}
\Big{(}3-4\Big{(}\frac{\alpha_{0}}{\alpha}\Big{)}^{4}\Big{)} \nonumber  \\
&-&4\Big{(}\frac{\tilde\mathcal{{E}}}{\alpha^{4}}\Big{)}^{2}
\Big{(}1-\Big{(}\frac{\alpha_{0}}{\alpha}\Big{)}^{4}\Big{)} -
\Big{(}\frac{\alpha_{0}}{\alpha}\Big{)} ^{4}
\Big{(}1+\frac{\tilde{\mathcal{E}}}{\alpha^{4}}\Big{)}^{1/2}
\Big{(}1+2\frac{\tilde{\mathcal{E}}}{\alpha^{4}}\Big{)} \Big{]}
\end{eqnarray}

As the brane crosses the bulk black hole horizon, the
CFT/FRW-cosmologies relations break down because $S_{H}\neq
S_{CFT}$ and $ kE_{BH}\neq E_{C}$.

 \section{A Probe Dp-Brane Moving in the Field of a  $
 \mathbf{Dp^{\prime}}$-Brane }

In this section we will generalize the motion of a probe Dp-brane
in the field of a $ Dp^{\prime}$-brane with $p^{\prime} \geq p $
\cite{kiritsis,kehagias}. In this case the $ Dp^{\prime}$-brane
metric is of the form
 \begin{equation}\label{in.metf}
ds^{2}_{10}=g_{00}(r)dt^{2}+g(r)(d\vec{x}_{p^{\prime}})^{2}+
  g_{rr}(r)dr^{2}+g_{S}(r)d\Omega_{8-p^{\prime}}.
\end{equation} In this background there exist in general a
non-trivial dilaton field and a RR $p^{\prime}+1$ form
$C_{p^{\prime}+1}$. The motion of the Dp-brane in this background
will be determined by the Dirac-Born-Infeld action given by
\begin{equation}\label{B.I. actionf}
  S_{p}=T_{p}~\int~d^{p+1}\xi
  e^{-\Phi}\sqrt{-det(\hat{G}_{\alpha\beta})}. \end{equation}
  Following a similar procedure as before the induced metric on
  the Dp-brane is $
  d\hat{s}^{2}=-d\eta^{2}+g(r(\eta))(d\vec{x})^{2}$ where the
  cosmic time is given by \begin{equation}\label{cosmicf}
 d\eta=\frac{|g_{00}|g^{\frac{P}{2}}e^{-\Phi}}{|C+E|}dt.
\end{equation}
The analogue of the p+1-dimensional Friedmann equations are
determined by defining the scale factor as $\alpha^{2}=g$. Then
the Friedmann equation we get is given by
\begin{equation}\label{denseff1f} \frac{8\pi G}{3}\rho_{eff}=
\frac{(C+E)^{2}g_{S}e^{2\Phi}-|g_{00}|(g_{S}g^{p}+\ell^{2}e^{2\Phi})}
{4|g_{00}|g_{rr}g_{S}g^{p}}\Big{(}\frac{g'}{g}\Big{)}^{2}.
\end{equation}

In this background there is also the possibility of having a
constant NS/NS two-form which lives in the world-volume of the $
Dp^{\prime}$-brane and it can be parametrized by $B=bdx^{p-1}
\wedge dx^{p}$. This constant $B$ field will not affect the
background field equation since it enters via its field strength
$H=dB$ which is zero for a constant $B$. However, the probe brane
will feel not H but directly the antisymmetric field $B_{\mu \nu}$
through the coupling
\begin{eqnarray}\label{B.I. actionb}
  S_{p}&=&T_{p}~\int~d^{p+1}\xi
  e^{-\Phi}\sqrt{-det(\hat{G}_{\alpha\beta}-
  \hat{B}_{\alpha\beta})}
  \\  \nonumber
  &+&T_{p}~\int~d^{p+1}\xi\hat{C}_{p+1}+anomaly~terms,
\end{eqnarray}
 where
\begin{equation}
  \hat{B}_{\alpha\beta}=B_{\mu\nu}\frac{\partial x^{\mu}\partial x^{\nu}}
  {\partial\xi^{\alpha}\partial\xi^{\beta}}.
\end{equation}
Then, the induced Friedmann equation on the brane (in case
$p=p^{\prime}$)  can be calculated to be
\begin{equation}\label{denseff1b} \frac{8\pi G}{3}\rho_{eff}=
\frac{(C+E)^{2}g_{S}e^{2\Phi}-|g_{00}|\Big{(}g_{S}g^{p-2}(g^{2}+b^{2})+\ell^{2}e^{2\Phi}\Big{)}}
{4|g_{00}|g_{rr}g_{S}g^{p-2}(g^{2}+b^{2})}\Big{(}\frac{g'}{g}\Big{)}^{2}.
\end{equation}

\subsection{The Holography Probed by a D3-Brane Moving in the Background Geometry of a
D3-Black Brane}

We will consider a probe D3-brane moving in the background
geometry of a near-extremal black hole with a metric \cite{horow}
\begin{equation}
ds^{2}_{10}=\frac{1}{\sqrt{H_{p^{\prime}}}}\Big{(}-f(r)dt^{2}+(d
\vec{x}) ^{2}\Big{)} +\sqrt{H_{p^{\prime}}}\,\frac{dr^{2}}{f(r)}+
\sqrt{H_{p^{\prime}}}\, r^{2}d \Omega ^{2}_{8-p^{\prime}},
\label{bhmetricdp}
\end{equation}
where $ H_{p^{\prime}}=1+\Big{(} \frac{L}{r}\Big{)}
^{7-p^{\prime}}$, \, and  $f(r)=1-\Big{(} \frac{r_{0}}{r}\Big{)}
^{7-p^{\prime}}$. In this background the RR form is
\begin{equation}
C_{012...p^{\prime}}=\sqrt{1+\Big{(} \frac{r_{0}}{L}\Big{)}
^{7-p^{\prime}}}\, \frac{1-H_{p^{\prime}}(r)}{H_{p^{\prime}}(r)},
\end{equation} and the dilaton field takes the form $ e^{\Phi}=H_{p^{\prime}}^
{(3-p^{\prime})/4}$. Taking the near horizon limit of the above
geometry, we recover the Schwarzschild-Ad$S_{5}\times S^{5}$ black
hole geometry discussed in the previous sections. There are also
spherically symmetric backgrounds, like the type-0 string
background, in which the Ad$S_{5}\times S^{5}$ geometry is
obtained only asymptotically. For this particular background,
depending on the value of the Tachyon field, the conformal
symmetry is restored in the infrared and in the ultraviolet, where
the Ad$S_{5}\times S^{5}$ geometry is recovered \cite{Typ0,Minah}.
The main motivation of our study in this section, is to follow the
motion of a probe D3-brane in these backgrounds where the
conformal invariance is broken and find how the thermodynamic
quantities change and what is their relation to geometrical
quantities. This study would be useful to determine the
cosmological evolution of a brane-universe moving between
conformal points \cite{future}.

Defining the parameter $\xi= \Big{(}1+\Big{(}
\frac{r_{0}}{L}\Big{)} ^{7-p^{\prime}}\Big{)} ^{1/2}$ the first
Friedmann equation (\ref{denseff1f}) in this background is
\begin{equation} \label{friedpp}
H^{2}=\frac{(7-p^{\prime})}{16L^{2}} \,
\alpha^{\frac{2(3-p^{\prime})}{(7-p^{\prime})}} (1-\alpha^{4})
^{\frac{2(8-p^{\prime})}{(7-p^{\prime})}}\Big{[} \frac{(E+\xi
\alpha^{4}-\xi)^{2}}{\alpha^{8}}-\Big{(}\xi^{2}-\frac{\xi^{2}-1}{\alpha^{4}}
\Big{)}
\Big{(}1+\frac{l^{2}}{L^{2}}\frac{(1-\alpha^{4})^{\frac{2}{(7-p^{\prime})}}}
{\alpha^{4+\frac{8}{(7-p^{\prime})}}} \Big{)}\Big{]}
\end{equation} We will consider the case of $p^{\prime}=3$. In
this case the above Friedmann equation becomes
\begin{equation} \label{friedp3}
H^{2}=\frac{1}{L^{2}} (1-\alpha^{4}) ^{5/2} \Big{[}
\Big{(}\xi+\frac{\hat{E}}{ \alpha^{4}}\Big{)}^{2}
-\Big{(}\frac{r_{0}}{L}\Big{)} ^{4}\frac{1}{\alpha_{0}^{4}}
\Big{(}1-\Big{(} \frac{\alpha_{0}}{\alpha}\Big{)} ^{4}\Big{)}
\Big{(}1+a(1-\alpha^{4})^{1/2} \Big{)}\Big{]}, \end{equation}
where $\hat{E}=E-\xi$ and $\alpha$ is given by
\begin{equation} \label{new1} \alpha=\Big{(}1+\Big{(}\frac{L}{r}
\Big{)}^{4} \Big{)} ^{-1/4}.
\end{equation}
The second Friedmann equation is
\begin{eqnarray}
\dot{H}&=&-\frac{2}{L^{2}}\Big{[}\frac{5}{2} \alpha^{4}
(1-\alpha^{4}) ^{3/2} \Big{[} \Big{(}\xi+\frac{\hat{E}}{
\alpha^{4}}\Big{)}^{2} -\Big{(}\frac{r_{0}}{L}\Big{)}
^{4}\frac{1}{\alpha_{0}^{4}} \Big{(}1-\Big{(}
\frac{\alpha_{0}}{\alpha}\Big{)} ^{4}\Big{)}
\Big{(}1+a(1-\alpha^{4})^{1/2} \Big{)}\Big{]}\nonumber \\
&+&(1-\alpha^{4}) ^{5/2}\Big{[} 2\Big{(}\xi+\frac{\hat{E}}{
\alpha^{4}}\Big{)}^{2} -2\xi  \Big{(}\xi+\frac{\hat{E}}{
\alpha^{4}}\Big{)} + \Big{(}\frac{r_{0}}{L}\Big{)}
^{4}\frac{1}{\alpha^{4}} \Big{(}1+a(1-\alpha^{4})^{1/2}
\Big{)}\nonumber \\ &-&\Big{(}\frac{r_{0}}{L}\Big{)}
^{4}\frac{1}{\alpha_{0}^{4}} \Big{(}1-\Big{(}
\frac{\alpha_{0}}{\alpha}\Big{)} ^{4}\Big{)}
\Big{(}1+a(1-\alpha^{4})^{-1/2} \Big{)}
+\frac{3}{2}(1-\alpha^{4})^{1/2}\Big{]}. \label{desecondfr}
\end{eqnarray}
The energy parameter $\hat{E}$ can be written as \begin{equation}
\hat{E}=\tilde{E}+\Big{(}\frac{r_{0}}{L}\Big{)}
^{4}-\Big{(}1+\Big{(}\frac{r_{0}}{L}\Big{)}
^{4}\Big{)}^{1/2}.\label{newenergy}
\end{equation} In the near horizon limit $r_{0}\ll L$, the two
Friedmann equations (\ref{friedp3}) and (\ref{desecondfr}) become
identical to (\ref{friedbha}) and (\ref{friedgsec}) respectively,
with $\hat{E}=\tilde{E}-1$ as can be seen from (\ref{newenergy}).
For simplicity we will consider the case of $a=0$. Calculating the
effective energy density and pressure from (\ref{friedp3}) and
(\ref{desecondfr}) and demanding to have a radiation dominated
brane-universe, from the equation of state
$w=p_{eff}/\rho_{eff}=1/3$ we get
\begin{equation} \frac{\hat{E}}{\alpha^{4}}=-\frac{5}{2} \xi \alpha^{4}
(1+\frac{3}{2}\alpha^{4}) ^{-1}\Big{[}1\pm \big{[}1-\frac{2}{5}
\Big{(} \frac{\alpha_{0}}{\alpha}\Big{)}
^{4}\frac{1}{\alpha^{4}}-\frac{3}{5}\Big{(}
\frac{\alpha_{0}}{\alpha}\Big{)} ^{4}\Big{]} ^{1/2}\Big{]}.
\end{equation}
The only real solution of this equation is $\hat{E}/\alpha^{4}=0$
for $r\rightarrow 0$. This is consistent with our previous
discussion, because in this limit we find a radiation dominated
brane-universe and then our results of a radially moving probe
D3-brane are recovered. To find corrections to our thermodynamic
quantities we will consider the case of $\hat{E}/\alpha^{4}=0$ but
with $\alpha_{0}<<1$. In this limit the first and second Friedmann
equations become
\begin{eqnarray}
H^{2}&=&\frac{1}{L^{2}} (1-\alpha^{4}) ^{5/2} \Big{[}\Big{(}
\frac{\alpha_{0}}{\alpha}\Big{)} ^{4}+\alpha_{0}^{4}\Big{]} \label{appfirst}\\
\dot{H}&=&-\frac{2}{L^{2}} (1-\alpha^{4}) ^{5/2} \Big{[}\Big{(}
\frac{\alpha_{0}}{\alpha}\Big{)} ^{4}+ \frac{5\alpha^{4}}
{2(1-\alpha^{4})}\Big{[}\Big{(}\frac{\alpha_{0}}{\alpha}\Big{)}
^{4}+\alpha_{0}^{4}\Big{]}\Big{]} \label{appsec}.
\end{eqnarray}
The equation of state can be calculated from the above expressions
\begin{equation}
p_{eff}=\frac{1}{3}\Big{(}\frac{1+9\alpha^{4}}{1-\alpha^{4}}\Big{)}\rho_{eff}.
\end{equation}
Therefore, as the probe D3-brane moves in the background of
D3-black brane, all sorts of Mirage matter is induced on the
brane-universe and when $r \rightarrow 0 $, the brane should pass
from a conformal point where the Schwarzschild-Ad$S_{5}\times
S^{5}$ black hole geometry is restored.

In this limit one can show that still the first and second
Friedmann equations (\ref{appfirst}) and (\ref{appsec}) can be
written as the cosmological Cardy-Verlinde and Smarr formulae with
the Hubble entropy, Bekenstein-Hawking energy and Hubble entropy
given by \begin{eqnarray} S_{H}&=&\frac{V}{2 G L} (1-\alpha^{4})
^{5/4}\Big{(}\frac{\alpha_{0}}{\alpha}\Big{)}
^{2}\Big{[}1+\alpha^{4}\Big{]}^{1/2}\\
E_{BH}&=&\frac{3V}{4 \pi G L^{2}\alpha^{2}}\\
T_{Hubble}&=&\frac{1}{\pi
L}\Big{(}\frac{\alpha_{0}}{\alpha}\Big{)}^{2} (1-\alpha^{4})
^{5/4} \Big{[}\frac{1+ \frac{5\alpha^{4}}
{2(1-\alpha^{4})}(1+\alpha^{4}) }
{\Big{(}1+\alpha^{4}\Big{)}^{1/2} } \Big{]}.
\end{eqnarray}

An observer in the bulk, measuring distances with the variable $r$
and time with the AdS time $t$, uses equation (\ref{hawtemp}) with
$h(r)$ given by
\begin{equation}
h(r)=\Big{(} 1-\Big{(} \frac{r_{0}}{r}\Big{)}
^{7-p^{\prime}}\Big{)} \Big{(}1+\Big{(} \frac{L}{r}\Big{)}
^{7-p^{\prime}} \Big{)}^{-1/2} \end{equation} and finds the
Hawking temperature
\begin{equation}
T_{H}=\frac{(7-p^{\prime})}{4
\pi}\frac{r_{0}^{\frac{(5-p^{\prime})}{2}}}
{\sqrt{r_{0}^{(7-p^{\prime})}+L^{( 7-p^{\prime})}}},
\end{equation} where for $ p^{\prime}=3$ becomes
\begin{equation}
T_{H}=\frac{1}{\pi}\frac{r_{0}}{\sqrt{r_{0}^{4}+L^{4}}}.
\label{haw3}
\end{equation} An observer on the brane, measures the scale factor
$\alpha$ of the brane-universe using the cosmic time $\eta$. The
cosmic time $\eta$ is related to the AdS time $t$ through the
relation (\ref{cosmicf}) which in this background becomes
\begin{equation} \label{rel2}
d\eta=\alpha \Big{(}1+\frac{\hat{E}}{
\alpha^{4}}\Big{)}^{-1}\Big{(}1- \Big{(}\frac{r_{0}}{r} \Big{)}
^{4} \Big{)} dt.
\end{equation}
In the limit we considered, the conformal factor is
\begin{equation}
\lim_{r\rightarrow\infty}\frac{dt}{d\eta}=\frac{1}{\alpha}.
\end{equation} Then using the AdS/CFT relation
$T_{CFT}=1/\alpha T_{H}$ the CFT temperature is
\begin{equation}
T_{CFT}=\frac{1}{\pi L} \Big{(}\frac{r_{0}}{r}\Big{)}
\Big{(}1+\Big{(}\frac{L}{r} \Big{)}^{4} \Big{)} ^{-1/4}.
\end{equation} The Casimir energy in this limit can be calculated
to be
\begin{equation}
E_{C}=\frac{3V}{8 \pi G
L^{2}}\Big{(}\frac{\alpha_{0}}{\alpha}\Big{)}^{4} \Big{[}
\Big{(}1+\Big{(}\frac{r}{L} \Big{)}^{4} \Big{)}
^{-3/2}\Big{(}1+\alpha_{0}^{4}\Big{)}\Big{(}\frac{4+6\alpha^{4}}{1-\alpha^{4}}\Big{)}
-4\Big{(}1+\Big{(}\frac{r}{L} \Big{)}^{4} \Big{)} ^{1/2}\Big{]}.
\end{equation}

The CFT/FRW-cosmologies relations are valid only when the probe
D3-brane passes from a conformal point ($r\rightarrow 0$) as
easily can be checked.

\subsection{A Probe D3-Brane Moving in the Background of Near-Horizon Geometry
with a Constant B-field}

In this section we will consider the motion of a probe D3-brane in
the near-horizon geometry of a D3-brane with a constant B-field.
Because we are mainly interested for the corrections of our
thermodynamics quantities which are due to the presence of the
B-field, we will take first the near-horizon limit of the metric
(\ref{bhmetricdp}) and then use (\ref{denseff1b}) to find the
Friedmann equations.

The near-horizon limit of (\ref{bhmetricdp}) is (\ref{bhmetric})
and in this background (\ref{denseff1b}) becomes
\begin{equation}
H^{2}=\frac{1}{L^{2}} \Big{[}
\Big{(}1+\frac{\hat{E}}{\alpha^{4}}\Big{)}^{2}\Big{(}1+\frac{b^{2}}{\alpha^{4}}
\Big{)}^{-1} -\Big{(}1-\Big{(}\frac{r_{0}}{L}\Big{)}^{4}
\frac{1}{\alpha^{4}}\Big{)} \Big{(}1+a
\Big{(}1+\frac{b^{2}}{\alpha^{4}} \Big{)}^{-1}\Big{)}
 \Big{]}. \label{effriedbhb} \end{equation}
We are mainly interested for the effect of the B-field so we take
$a=0$ and then the above equation can be rewritten
\begin{equation} H^{2}=\frac{1}{L^{2}} \Big{[}
\Big{(}\frac{\alpha_{0}}{\alpha}\Big{)}^{4}+2\frac{\hat{E}}{
\alpha^{4}} + \frac{\hat{E}^{2}}{
\alpha^{8}}-\frac{b^{2}}{\alpha^{4}}\Big{(}1+\frac{\hat{E}}{
\alpha^{4}}\Big{)}^{2}\Big{(}1+\frac{b^{2}}{\alpha^{4}}
\Big{)}^{-1}\Big{]}. \label{bfirstfried} \end{equation} The
presence of the B-field in (\ref{bfirstfried}) acts effectively as
a radiation term on the brane-universe like the electric field we
already considered. The second Friedmann equation in this
background becomes
\begin{equation}
\dot{H}=-\frac{1}{L^{2}} \Big{[}
\Big{(}\frac{\alpha_{0}}{\alpha}\Big{)}^{4}+2\frac{\hat{E}}{
\alpha^{4}} + 2\frac{\hat{E}^{2}}{
\alpha^{8}}-\frac{b^{2}}{\alpha^{4}}\Big{(}1+\frac{b^{2}}{\alpha^{4}}
\Big{)}^{-2}\Big{[}1+ 4\frac{\hat{E}}{ \alpha^{4}} +
3\frac{\hat{E}^{2}}{
\alpha^{8}}+2\frac{b^{2}}{\alpha^{4}}\frac{\hat{E}}{ \alpha^{4}}
+2\frac{b^{2}}{\alpha^{4}}\frac{\hat{E}^{2}}{ \alpha^{8}}
\Big{]}\Big{]}. \label{bsecfried} \end{equation} From
(\ref{bfirstfried}) and (\ref{bsecfried}) the $\rho_{eff}$ and
$p_{eff}$ can be calculated and the equation of state specified.
Demanding to have a radiation dominated universe, the energy
parameter is fixed to $\hat{E}=b^{2}$. Using this value for
$\hat{E}$, the two Friedmann equations and the effective energy
density and pressure on the probe brane become
\begin{eqnarray}
H^{2}&=& \frac{1}{L^{2}}\Big{[}\Big{(}
\frac{\alpha_{0}}{\alpha}\Big{)} ^{4}+\frac{b^{2}}{\alpha^{4}}\Big{]} \label{bradiauniff} \\
\dot{H}&=& -\frac{2}{L^{2}}\Big{[}\Big{(}
\frac{\alpha_{0}}{\alpha}\Big{)} ^{4}+\frac{b^{2}}{\alpha^{4}}\Big{]} \label{bradiauniss} \\
\rho_{eff}&=& \frac{3}{8 \pi G L^{2}}\Big{[}\Big{(}
\frac{\alpha_{0}}{\alpha}\Big{)} ^{4}+2\frac{b^{2}}
{\alpha^{4}}+\frac{b^{4}}{\alpha^{8}}-\frac{b^{2}}{\alpha^{4}}\
\Big{(}1+\frac{b^{2}}{\alpha^{4}}\Big{)}\Big{]} \label{bradiaunirho} \\
p_{eff}&=& \frac{1}{8 \pi G L^{2}}\Big{[}\Big{(}
\frac{\alpha_{0}}{\alpha}\Big{)} ^{4}+2\frac{b^{2}}
{\alpha^{4}}+5\frac{b^{4}}{\alpha^{8}} -\frac{b^{2}}{\alpha^{4}}\
\Big{(}1+\frac{b^{2}}{\alpha^{4}}\Big{)}^{-2}\Big{(}1+
7\frac{b^{2}} {\alpha^{4}}+11\frac{b^{4}}{\alpha^{8}}+5
\frac{b^{6}} {\alpha^{12}}\Big{)}\Big{]} \label{bradiaunip}
\end{eqnarray} Observe that the two Friedmann equations (\ref{bradiauniff})
and (\ref{bradiauniss}) get a correction in first order in $b^{2}$
like the contribution they get from the electric field (in the
case $\tilde{E}=0$ ), while the effective energy density and
pressure are receiving high order corrections in $b^{2}$. The
Hubble entropy and the Hubble temperature, using
(\ref{bradiauniff})-(\ref{bradiaunip}), in this background become
\begin{eqnarray}
S_{H}&=& \frac{1}{2 G L }\Big{(} \frac{\alpha_{0}}{\alpha}\Big{)}
^{2}\Big{(}1+\Big{(}\frac{\alpha_{0}}{\alpha}\Big{)}
^{4}\frac{b^{2}}{\alpha^{4}}\Big{)}^{1/2}  \label{bhul}\\
T_{Hubble}&=&\frac{1}{\pi L}\Big{(}
\frac{\alpha_{0}}{\alpha}\Big{)}
^{2}\Big{(}1+\Big{(}\frac{\alpha_{0}}{\alpha}\Big{)}
^{4}\frac{b^{2}}{\alpha^{4}}\Big{)}^{1/2}, \label{bthub}
\end{eqnarray} while the Bekenstein Hawking temperature is
unchanged and is given by (\ref{hubenergy}). It is straitforward
to show that the two Friedmann equations can be written in terms
of (\ref{bhul}) and (\ref{bthub}) and $E_{BH}$, as the
cosmological Cardy-Verlinde and Smarr formulae.

To calculate the thermodynamic quantities we have to find the
conformal factor for this background. The conformal time is
\begin{equation}
d\eta=\alpha \Big{(}1-\Big{(}\frac{r_{0}}{r}\Big{)} ^{4}\Big{)}
\Big{(}1+\frac{b^{2}}{\alpha^{4}} \Big{)}^{-1/2}dt,
\end{equation} from where the conformal factor is
\begin{equation}
\lim_{r\rightarrow\infty}\frac{dt}{d\eta}=\frac{1}{\alpha}\Big{(}1+\frac{b^{2}}{\alpha^{4}}
\Big{)}^{1/2}. \end{equation} Using the Hawking temperature
$T_{H}=r_{0}/\pi L^{2}$ the CFT temperature according to AdS/CFT
correspondence is
\begin{eqnarray} T_{CFT}&=&\frac{1}{\alpha}\Big{(}1+\frac{b^{2}}{\alpha^{4}}
\Big{)}^{1/2}T_{H}\nonumber \\
&=&\frac{1}{\pi L}\Big{(}
\frac{\alpha_{0}}{\alpha}\Big{)}\Big{(}1+\frac{b^{2}}{\alpha^{4}}\Big{)}^{1/2}.
\label{btcft}\end{eqnarray} The CFT entropy is unchanged and it is
given by (\ref{cftentropdens}), while the Casimir energy becomes
 \begin{equation}
E_{C}=\frac{3V}{2 \pi G L^{2}}\Big{[}\Big{(}
\frac{\alpha_{0}}{\alpha}\Big{)}
^{4}+\frac{b^{2}}{\alpha^{4}}-\Big{(}\frac{\alpha_{0}}{\alpha}\Big{)}
^{4}\Big{(}1+\frac{b^{2}}{\alpha^{4}}\Big{)}^{1/2}\Big{]}.
\end{equation}

At the moment the probe brane crosses the bulk black hole horizon,
from (\ref{bthub}) and (\ref{btcft}) we get
\begin{equation} T_{CFT}=T_{Hubble}=\frac{1}{\pi
L}\Big{(}1+\frac{b^{2}}{\alpha_{0}^{4}}\Big{)}^{1/2}.
\end{equation} We also have \begin{eqnarray} S_{H}&=&
\frac{1}{2 G
L}\Big{(}1+\frac{b^{2}}{\alpha_{0}^{4}}\Big{)}^{1/2}\neq
S_{CFT}=\frac{1}{2 G L}\\ E_{C}&=&\frac{3V}{2 \pi G L^{2}}\Big{[}1
+\frac{b^{2}}{\alpha_{0}^{4}}-\Big{(}1+\frac{b^{2}}{\alpha_{0}^{4}}\Big{)}^{1/2}\Big{]}
\neq k E_{BH}=0. \end{eqnarray} Therefore, in the near horizon
limit with a constant B-field in the background, there is no
CFT/FRW-cosmologies correspondence.

%Tetarto Kefalaio me Onoma: Conclusions
\section{Conclusions}

We studied the cosmological holographic principle in a generic
static spherically symmetric background with a probe D3-brane
moving in this background playing  the r$\hat{o}$le of the
boundary to this space. After reviewing the necessary formalism,
we followed the motion of the probe D3-brane in specific
background geometries under different initial conditions of the
probe D3-brane.

First we followed the motion of the probe D3-brane in the
near-horizon $AdS_{5}\times S^{5}$ background with a
Schwarzschild-Ad$S_{5}$ black hole \cite{kiritsis}. Demanding to
have a radiation dominated universe on the probe D3-brane, the
energy parameter is fixed and for zero angular momentum we showed
that the Cardy-Verlinde and Smarr formulae are equivalent to the
first and second Friedmann equations respectively. Using the
AdS/CFT correspondence we related the entropy and the Hawking
temperature of the AdS$_{5}$ black hole with the entropy and
temperature of the CFT on the probe D3-brane. At the special
moment when the probe D3-brane crosses the horizon of the
background black hole, an observer on the brane measures the CFT
entropy and temperature using pure geometrical quantities, the
Hubble parameter and its time derivative. Furthermore, at that
particular moment, we showed that the CFT entropy is described by
the Cardy-Verlinde formula, with zero Casimir energy. These
results indicate that, as the D3-brane crosses the horizon of the
background black hole, it probes the holography of the dual CFT
theory of the $AdS_{5}\times S^{5}$ geometry and these results are
in agreement with \cite{verlide,savonije}.

Next we followed the motion of a probe D3-brane having a non-zero
angular momentum, in the same near-horizon $AdS_{5}\times S^{5}$
background. We showed that the AdS/CFT correspondence is exact in
the sense that the CFT and Hawking temperature are related in the
same way as in the case of zero angular momentum. The presence of
a non-zero angular momentum on the brane, modifies some of the
thermodynamic quantities, and on the horizon the $T_{CFT}$ cannot
anymore be expressed in terms of the Hubble parameter and its time
derivative. The modification of the second Friedmann equation
results in a non-zero Casimir energy which has the consequence
that the CFT entropy on the horizon is not described by the
Cardy-Verlinde formula as it happens to the zero angular momentum
case.

An electric field was also considered on the probe D3-brane. Its
presence introduces another energy scale which is proportional to
its field strength. It was shown that the electric field acts
effectively as a new radiation term on the probe brane. Demanding
to have a radiation dominated universe, the energy parameters were
fixed and then we calculated the corrections to the thermodynamic
quantities due to the electric field. We showed that as the brane
crosses the bulk black hole horizon it is no possible any more to
relate thermodynamic with geometrical quantities through the
CFT/FRW-cosmologies relations. Finally, we considered a more
general problem of a probe D3-brane moving in the field of other
D3-branes. In the general case, the AdS/CFT correspondence and the
CFT/FRW relations break down as expected, but the previous results
can be recovered in the near-horizon limit.

It would be interesting to investigate the AdS/CFT correspondence
and the CFT/FRW relations in the case of a non-zero $\rho^{2}$
term on the brane. This is the term generated in early
cosmological evolution in the Randall-Sundrum model. We know that
the trace anomaly of the energy momentum tensor is proportional to
this $\rho^{2}$ term \cite{ida}. Then one expects to have a broken
conformal theory \cite{kanno}, and it would be interesting to see
what kind of holographic description it is possible for such a
theory which has a trace anomaly \cite{casadio,padilla1,otha1}.

Another interesting line of investigation is to apply these ideas
and formalism to a realistic inflationary brane model. The AdS/CFT
correspondence for example can help us to solve the problem of
exit from inflation, because the moment the moving brane-universe
crosses the horizon of the bulk black hole, it enters a thermal
bath and the reheating process of the universe can start. Actually
such a model can be realized for a particular background geometry
\cite{future}.

\section{Acknowlegements}

We thank A. Kehagias for valuable discussions and comments. Work
partially supported by NTUA research program "Thales" and by the
Greek Ministry of Education program "Hraklitos".

\end{document}